\documentclass[aps,prd,a4paper,eqsecnum,%
superscriptaddress,nofootinbib,showpacs,%
twocolumn,showkeys,amsfonts,amssymb,amsmath,epsfig]{revtex4}

\usepackage{amsthm}
\usepackage{graphicx}

\newcommand{\be}{\begin{equation}}
\newcommand{\ee}{\end{equation}}

\input xy
\xyoption{all}

\begin{document}

\title{Dynamics of gravitational field within a wave
front and thermodynamics of black holes}

\author{ Ewa Czuchry}
\affiliation{Yukawa Institute for Theoretical Physics, Kyoto
University,
\\ Kitashirakawa-Oiwake-Cho, Sakyo-ku, Kyoto 606-8502,
Japan}

\author{Jacek Jezierski}
\affiliation{Department of Mathematical Methods in Physics, \\
University of Warsaw, ul. Ho\.za 69, 00-682 Warszawa, Poland}

\author{ Jerzy Kijowski}
\affiliation{Center for Theoretical Physics, Polish Academy of Sciences, \\
 Al. Lotnik\'ow 32/46; 02-668
Warszawa, Poland}

\begin{abstract} Hamiltonian dynamics of gravitational field contained in a
spacetime region with boundary $S$ being a null-like hypersurface
(a wave front) is discussed. Complete Hamiltonian formula for the
dynamics (with no surface integrals neglected) is derived. A
quasi-local proof of the first law of black holes thermodynamics
is obtained as a consequence, in case when $S$ is a non-expanding
horizon. The zeroth law and Penrose inequalities are discussed
from this point of view.
\end{abstract}



\maketitle

\section{Introduction}

Evolution of gravitational field within a finite tube with a {\em
timelike boundary} was derived using a consistent, Hamiltonian
formulation in \cite{pieszy} and then reformulated in \cite{K1}.
Here we extend this description to the case of a {\em null-like
boundary} or a {\em wave front}, i.e.,~a three-dimensional
submanifold whose internal metric is degenerate. Restricting our
result to the special case of wave fronts, namely to non-expanding
horizons, we obtain a generalization of the first law of
thermodynamics for black holes as a simple consequence.

Contrary to the Iyer and Wald approach (see \cite{Wald}), no
assumption about stationarity (existence of a Killing field) is
necessary here. Such an assumption finally reduces our formula to
the standard first law.

In many presentations of the Hamiltonian field theory
(cf.~\cite{Bogol-Sh}) boundary problems are neglected.
Consequently, all the surface integrals arising from the
integration by parts are assumed to vanish. Here we use the
formulation proposed in \cite{Kij-Tulcz} and \cite{springer},
where the field boundary data are treated on the same footing as
the Cauchy data. This is the only way to obtain a mathematically
consistent (infinite dimensional) Hamiltonian description of any
field theory if the boundary of the space volume taken into
account is non-trivial.

To illustrate our approach, consider as an example the linear
theory of an elastic, finite string. Field configuration of the
string is described by its displacement function: ${\mathbb R}
\times [a , b ] \ni (t,x) \rightarrow \varphi(t,x) \in {\mathbb
R},$ fulfilling the wave equation:
\begin{equation}\label{wave}
    \frac 1{c^2} \frac {\partial^2}{\partial t^2}\ \varphi =
    \frac {\partial^2}{\partial x^2}\ \varphi .
\end{equation}
Here, velocity ``$c$'' is a combination of the string's proper
density (per unit length) and its elasticity coefficient. Passing
to appropriate time and length units, we may always put $c=1$.
Equation (\ref{wave}) may be derived from the Lagrangian density
\begin{align}\label{Lagr}
    L =& - \frac 12 \sqrt{|\det g |} \  g^{\mu\nu}
    (\partial_\mu \varphi) (\partial_\nu \varphi)\nonumber\\
    = & \frac 12  \left(
    (\dot\varphi )^2 - (\varphi^\prime)^2 \right)
    ,
\end{align}
where $\mu,\nu = 0,1$ and $(x^0,x^1)=(t,x)$, $g_{\mu\nu}=$
diag$(-1,+1)$, ``dot'' denotes the time derivative and ``prime''
denotes the space derivative. A convenient way to encode the
entire information about the dynamics of the string is to write it
in a form of the following equation:
\begin{equation}\label{gen-Lagr-true}
  \delta L(\varphi , \partial_\nu \varphi ) =
  \partial_\mu \left( p^\mu \delta \varphi \right)=
  \left( \partial_\mu p^\mu \right) \delta \varphi +
  p^\mu \delta \left( \partial_\mu \varphi \right) ,
\end{equation}
equivalent to the Euler-Lagrange equations. Indeed, condition
\eqref{gen-Lagr-true} is satisfied if and only if the volume part
of the variation of $L$ (normally present on the right-hand side)
vanishes. The above equation (which we refer to, as the {\em
Lagrangian generating formula} of the dynamics) has a beautiful
symplectic interpretation (see \cite{Kij-Tulcz} or
\cite{springer}) as a definition of a Lagrangian submanifold of
{\em physically admissible states} within the symplectic space of
first jets of sections of the state bundle. Here, $(\varphi ,
\partial_\nu \varphi , \partial_\mu p^\mu , p^\nu)$ are
local canonical coordinates in this symplectic space. Without
going into deep, mathematical details, formula
\eqref{gen-Lagr-true} may be simply read as the following
condition imposed on the string configuration: the ``response
parameters'' $(\partial_\mu p^\mu , p^\nu )$ must be equal to
partial derivatives of the Lagrangian $L$ with respect to the
corresponding ``control parameters'' $(\varphi ,
\partial_\nu \varphi )$. Hence, we have the following dynamical
equations of the theory:
\begin{enumerate}
\item  definition of the canonical momenta:
\begin{itemize}
\item  kinetic momentum
\[
   p^0 = \frac {\partial L}{\partial (\partial_0\varphi)}
   =  \partial_0\varphi = \dot\varphi ,
\]
\item  stress density
\[
   p^1 = \frac
   {\partial L}{\partial (\partial_1\varphi)}
   = -\partial_1\varphi = - \varphi^\prime ,
\]
\end{itemize}
\item  Euler--Lagrange equation, equivalent to (\ref{wave}):
\[
   \partial_\mu p^\mu =
   \frac {\partial L}{\partial \varphi}  = 0  \ .
\]
\end{enumerate}
For purposes of the Hamiltonian description of the theory we
introduce the following notation:
\begin{equation}\label{notation}
    \pi := p^0 \ ; \ \ \ \ \ \pi^\perp := p^1 .
\end{equation}
Integrating {\em infinitesimal} generating formula
(\ref{gen-Lagr-true}) over the entire string $[a,b]$ we obtain the
spatially {\em finite} generating formula (it is still {\em
infinitesimal} with respect to the time variable):
\begin{equation}\label{finite-L}
    \delta \int_a^b L = \int_a^b \left( \dot\pi \delta\varphi
    + \pi \delta\dot\varphi \right) +
    \left[
    \pi^\perp \delta\varphi \right]^b_a .
\end{equation}
Hamiltonian description of the same dynamics is obtained {\em via}
Legendre transformation between $\pi$ and $\dot\varphi$, putting
$\pi \delta\dot\varphi = \delta (\pi \dot\varphi)- \dot\varphi
\delta \pi$:
\begin{equation}\label{finite}
    -\delta {\cal H} =
    \int_a^b \left( \dot\pi \delta\varphi
    - \dot\varphi \delta \pi \right) +
    \left[
    \pi^\perp \delta\varphi \right]^b_a ,
\end{equation}
with
\begin{eqnarray}\label{Hamilt-H}
  {\cal H}:=\int_a^b (\pi\dot\varphi - L)
  = \frac 12 \int_a^b (\pi^2 + (\varphi^\prime)^2)  .
\end{eqnarray}
Equation (\ref{finite}) acquires an infinitely-dimensional,
Hamiltonian meaning:
\begin{eqnarray}\label{Hamilt-fi}
  {\dot \pi}  =  - \frac {\delta {\cal H}}{\delta\varphi} , \ \
  \ {\dot \varphi}  =  \frac {\delta {\cal H}}{\delta
  \pi},
\end{eqnarray}
as soon as the boundary term in \eqref{finite} is killed. This may
be done in many ways, by imposing appropriate boundary conditions.
Physically, this corresponds to a choice of a specific device
controlling the behaviour of the extremal points of the string.
Mathematically, such a choice implies a specific self adjoint
extension of the second order differential operator on the
right-hand side of the dynamical equation \eqref{wave}. This makes
the dynamics uniquely defined. As an example consider two such
choices: Dirichlet conditions and the Neumann conditions. In the
Dirichlet mode we restrict ourselves to an infinitely dimensional
phase space of initial data $( \varphi , \pi )$, defined on
$[a,b]$ and fulfilling conditions: $\varphi (a) \equiv A$,
$\varphi (b) \equiv B$. Within this phase space we have
$\delta\varphi(a)=\delta\varphi(a)=0$ and equations
(\ref{Hamilt-fi}) hold.

Instead of controlling the string configuration $\varphi$ at the
boundary, we may control its stress by applying an appropriate
force $F$. This leads to the Neumann control mode: $\pi^\perp(a) =
F_{\rm left}$, $\pi^\perp(b) = F_{\rm right}$. Consequently,
$\delta\pi^\perp$ vanishes at the boundary. Performing Legendre
transformation between $\varphi$ and $\pi^\perp$ at the boundary:
$\pi^\perp \delta\varphi = \delta (\pi^\perp\varphi)- \varphi
\delta \pi^\perp$, we obtain again a legitime Hamiltonian system,
defined in a phase space which is completely different from the
previous one and with a new Hamiltonian ${\widetilde{\cal H}}$
playing role of free energy:
\begin{equation}\label{finite-Neu}
    -\delta {\widetilde{\cal H}} =
    \int_a^b \left( \dot\pi \ \delta\varphi
    - \dot\varphi \ \delta \pi \right) -
    \left[
    \varphi \ \delta \pi^\perp \right]^b_a ,
\end{equation}
where
\begin{align}
\label{Hamilt-H-tilde}
  {\widetilde{\cal H}} :=& {\cal H} + \left[
    \varphi \  \pi^\perp \right]^b_a =
    {\cal H} - \left[
    \varphi \  \varphi^\prime \right]^b_a\nonumber\\
  =& \frac 12 \int_a^b (\pi^2 - (\varphi^\prime)^2
  - 2 \varphi\varphi^{\prime\prime})  .
\end{align}
Again, boundary term in (\ref{finite-Neu}) vanishes due to Neumann
conditions and the field dynamics reduces to (\ref{Hamilt-fi}).

Consider now the subspace of static solutions: ${\dot \pi} = 0 =
{\dot \varphi}$. Due to (\ref{Hamilt-fi}), the functional
derivative of the $\cal H$ vanishes at those points and, whence,
Hamiltonian formula (\ref{finite}) describes only the virtual work
performed by the configuration controlling device at the boundary:
\begin{equation}\label{virt-work}
    \delta {\cal H} =
    - \left[
    \pi^\perp \delta\varphi \right]^b_a .
\end{equation}
But, due to (\ref{Hamilt-H}), $\cal H$ is manifestly convex. This
implies that every static solution corresponds to the minimal
value of the Hamiltonian in the phase space defined by the
Dirichlet conditions. Due to equation (\ref{wave}) and to boundary
conditions, such a solution is given by: $\pi \equiv 0$ and
$\varphi(x)= A + (x-a)\tfrac {B-A}{b-a}$. Inserting this value
into (\ref{Hamilt-H}) we obtain the following ``Penrose-like
inequality'':
\begin{equation}\label{ineq}
    \frac {(B-A)^2}{b-a} \le {\cal H} ,
\end{equation}
analogous to the gravitational Penrose inequality relating the
energy carried by Cauchy data outside of a horizon $S$ and the
energy of a black hole corresponding to the same value of
appropriate boundary data on $S$.

In the Neumann mode, the new Hamiltonian ${\widetilde{\cal H}}$ is
obviously non-convex. It is easy to check, that the free energy
(\ref{Hamilt-H-tilde}) is unbounded neither from below nor from
above and possesses no stationary points as soon as $F_{\rm left}-
F_{\rm right} \ne 0$. There is, therefore, no ``Penrose-like''
inequality in this mode.

In \cite{pieszy} the dynamics of the gravitational field within a
timelike world tube $S$ was analyzed in a similar way. For this
purpose the so called ``affine variational principle'' was used,
where the Lagrangian function depends on the Ricci tensor of a
spacetime connection $\Gamma$. In this picture, the metric tensor
$g$ arises only in the Hamiltonian formulation as the momentum
canonically conjugate to $\Gamma$. Later, it was proved in
\cite{K1} that the Hamiltonian dynamics obtained this way is
universal and does not depend upon a specific variational
formulation we start with (actually, it can be derived from field
equations only, without any use of variational principles, the
existence of them being a consequence of the ``reciprocity'' of
Einstein equations -- see \cite{Kij-Tulcz} and \cite{springer}).
On the contrary, the Hamiltonian picture is very sensitive to the
method of controlling the boundary data. A list of natural control
modes, leading to different ``quasilocal Hamiltonians'', is given
in \cite{K1}. Each of them is related with a specific choice of
control variables at the boundary. The ``true mass'', which tends
to the ADM mass when shifting the boundary to infinity, is one of
them (see also an analysis of the linearized theory
\cite{JJGRG31}).

The aim of the present paper is to generalize above results to the
case when the boundary $S$ is a wave front (a three-dimensional
submanifold of spacetime $M$ whose internal three-metric $g_{ab}$
is degenerate). This way we obtain a general Hamiltonian formula
for the gravitational field dynamics within (or outside) a wave
front, which is very much analogous to formula \eqref{finite} for
the string theory. As a byproduct, assuming that the wave front
$S$ is very special, namely is a non-expanding horizon, we obtain
a generalization of the first law of thermodynamics for black
holes (see formula \eqref{form-zerowa3}). Some of results have
been already published in \cite{Bl-H}.

\section{Dynamics of the gravitational field within a null
hypersurface}

Consider gravitational field dynamics inside a null hypersurface
$S$:

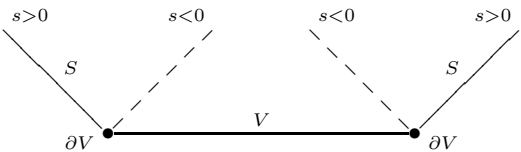
\[
\xymatrix@M=0pt@R=1.3cm@C=1.3cm{ \ar@{-}[dr]_>{\partial V
}^<{s>0}^S & &\ar@{--}[dl]_<{s<0}&\ar@{--}[dr]^<{s<0}&&
\ar@{-}[dl]^>{\partial V}_<{s>0}_S\\
&{\bullet}\ar@{-}[r]&\ar@{-}[r]^V&\ar@{-}[r]&{\bullet}&}\]

Parameter $s = \pm 1$ labels two possible situations: an expanding
or a shrinking wave front (if $S$ is a horizon, these correspond
to a black hole or a white hole case). To simplify notation we use
coordinates $x^\mu$, $\mu=0,1,2,3$, adapted to the above
situation: $x^0 = t$ is constant on a chosen family of (spacelike)
Cauchy surfaces $\Sigma_t$ whereas $x^3$ is constant on the
boundary $S$. This does not mean that $x^3$ is null-like
everywhere, but only on $S$. We may imagine that $x^3=r$
coincides, far away from $S$, with the spacelike radial
coordinate. Consider the three-dimensional volume $V \subset
\Sigma_{t_0}$ defined as the set of those points of $\Sigma_{t_0}$
which are situated {\em inside} $S$. Coordinates $x^A$, $A=1,2$,
are ``angular'' coordinates on the two-surface $\partial V =V\cap
S$ whose topology is assumed to be that of a two-sphere. Finally,
$x^k$, $k=1,2,3$, are spatial coordinates on the Cauchy surfaces
$\{ x^0 =$ const.$\}$ and $x^a$, $a=0,1,2$, are coordinates on
$S$. We stress, however, that our results are coordinate
independent and will be expressed in terms of relations between
geometric objects defined on $V$ and $S$.

In Appendix \ref{inside} we prove the following identity fulfilled
by any one-parameter family of solutions of Einstein equations
(``variation'' operator $\delta$ may be understood as a derivative
with respect to this parameter and ``dot'' denotes the time
derivative):
\begin{align}
- \delta {\cal H}  =&  \frac 1{16 \pi} \int_V  \left( {\dot
P}^{kl} \delta g_{kl} - {\dot g}_{kl} \delta P^{kl}
\right)\nonumber\\ &+ \frac s{8 \pi} \int_{\partial V} ( {\dot
\lambda} \delta  a  -
{\dot  a} \delta \lambda ) \nonumber \\
  +  \frac s{16 \pi}&
\int_{\partial V} \left( \lambda {l}^{AB}\delta g_{AB}
 -2\left(w_0\delta\Lambda^0-\Lambda^A\delta w_A
\right)\right), \label{form-zerowa1a}
\end{align}
where
\begin{equation}
  {\cal H}  =   \frac 1{8 \pi} \int_V {\cal G}^0_{\ 0} + \frac s{8
   \pi} \int_{\partial V}  \lambda l
  \equiv   \frac s{8 \pi}
  \int_{\partial V}  \lambda l
  \, , \label{H-grav-volume1}
\end{equation}
and $P^{kl}$ denotes external curvature of the Cauchy surface,
written in the ADM form  (cf. \cite{ADM}). Moreover,
$\lambda=\sqrt{\det g_{AB}}$ is the two-dimensional volume form on
${\partial V}$ and $a=-\frac12\log |g^{00}|= \log N$, where $N$ is
the lapse function. To define the remaining objects we must choose
a null (time oriented) field $K$ tangent to $S$. It is not unique,
since $fK$ is also a null (time oriented) field for any (positive)
function $f$ on $S$. For purposes of the Hamiltonian formula
(\ref{form-zerowa1a}) we always choose the normalization
compatible with the (3+1)-decomposition used here: $<K,dx^0>\;
=1$. Hence, $K=\partial_0 - n^A \partial_A$. On the contrary, the
vector-density $\Lambda^a = \lambda K^a =(\lambda,-\lambda n^A)$
is uniquely defined on $S$ and does not depend upon a particular
choice of $K$. Now, we define\footnote{This is the
(2+1)-decomposition of the extrinsic curvature $Q^a{_b}(K)$
defined in \cite{JKC1} and \cite{jjrw}.}
\begin{align}\label{el-def}
l_{ab}:=&  -g(\partial_b,\nabla_a K) = - \frac 12 \pounds_K g_{ab}
, \\\label{wu-def} w_a :=& -<\nabla_a K , dx^0> ,
\end{align}
where $g_{ab}$ is the induced (degenerate) metric on $S$. Because
of the identity: $l_{ab}K^a = 0$, the null mean curvature:
$l=\tilde{\tilde{g}}^{AB}l_{AB}$ may be defined (it is often
denoted by $\theta$ -- see \cite{Hawking}, \cite{jjrw}), where by
$\tilde{\tilde{g}}^{AB}$ we denote the inverse two-metric.

The volume term in (\ref{H-grav-volume1}) vanishes due to
constraint equations\footnote{In the presence of matter the volume
term equals ${\cal G}^0_{\ 0}-8\pi T^0_{\ 0}$ and also vanishes
due to constraint equations.}  ${\cal G}^0_{\ \nu}=0$. ${\cal
G}^0_{\ 0}$ is often denoted by $N{ H}+N^k{ H}_k$ (see
e.g.,~\cite{Misner}), where $ H$ is the scalar (``Hamiltonian'')
constraint and ${ H}_k$ are the vector (``momentum'') constraints,
$N$ and $N^k$ are the lapse and the shift functions. Constraint
equations ${ H}=0$ and ${ H}_k=0$ imply vanishing of ${\cal
G}^0_{\ 0}$.

In the Appendices B and C we give two independent proofs of the
identity (\ref{form-zerowa1a}). The first one is analogous to the
transition from formula (\ref{gen-Lagr-true}) to formula
(\ref{finite}). For this purpose we use Einstein equations written
analogously to (\ref{gen-Lagr-true}) (cf.~\cite{pieszy}):
\begin{equation}\label{gen-Lang-grav}
  \delta L = \partial_\kappa \left(
  \pi^{\mu\nu}\delta
   A^\kappa_{\mu\nu} \right)
 ,
\end{equation}
where $\displaystyle {\pi}^{\mu\nu} := \tfrac 1{16 \pi} \sqrt{|g|}
\ g^{\mu\nu}$, and $\displaystyle A^{\lambda}_{\mu\nu} :=
{\Gamma}^{\lambda}_{\mu\nu} - {\delta}^{\lambda}_{(\mu}
{\Gamma}^{\kappa}_{\nu ) \kappa}$. Integrating
(\ref{gen-Lang-grav}) over a volume  $V$ and using metric
constraints for the connection $\Gamma$, we directly prove
(\ref{form-zerowa1a}).

However, in Appendix C, an indirect proof is also provided, based
on a limiting procedure, when a family $S_\epsilon$ of timelike
surfaces tends to a lightlike surface $S$. It is shown that the
non-degenerate formula derived in \cite{pieszy} and \cite{K1}
gives (\ref{form-zerowa1a}) as a limiting case for $\epsilon
\rightarrow 0$.

The last term in (\ref{form-zerowa1a}) may be written in the
following way:
\[
-\Lambda^A\delta w_A =\lambda n^A \delta w_A =   n^A \delta {\cal
W}_A - n^A w_A \delta \lambda ,
\]
where ${\cal W}_A := \lambda w_A$ and
$n^A:=\tilde{\tilde{g}}^{AB}g_{0B}$.
Denoting  $\kappa :=  n^A w_A -w_0  = - K^a w_a$ we finally obtain
the following generating formula:
\begin{subequations}\label{17all}
\begin{align}
   - \delta {\cal H} & =  \frac 1{16 \pi} \int_V  \left( {\dot
   P}^{kl}  \delta g_{kl} - {\dot g}_{kl} \delta P^{kl} \right)\label{17a}\\& +
   \frac s{8 \pi} \int_{\partial V} ( {\dot \lambda} \delta  a  -
   {\dot  a} \delta \lambda )
   \label{17}
   \\
   & +  \frac s{16 \pi}
  \int_{\partial V} \left( \lambda {l}^{AB}\delta g_{AB}
  +2\left(\kappa\delta\lambda - n^A\delta {\cal
  W}_A \right)\right). \label{form-zerowa1}
\end{align}
\end{subequations}
It is easy to prove that the integral lines of $K$ are null
geodesics. This means that $K^a\nabla_a K$ is  always proportional
to $K$. Hence, quantity $\kappa$ (traditionally called a ``surface
gravity'' on $S$) fulfills equation $K^a\nabla_a K= \kappa K$,
which may be used as its alternative definition. We stress that
its value {\em does not} correspond to any intrinsic property of
the surface $S$, but depends upon a choice of the null field $K$
on $S$ (i.e.,~upon a (3+1)-decomposition of spacetime). However,
in a special case of a black hole thermodynamics, there is a
privileged choice of $K$, compatible with the Killing field of the
stationary solution and its normalization to unity at infinity. In
this case $\kappa$ is an intrinsic property of the hole and the
above formula provides, as will be seen later, the so called first
law of black hole thermodynamics.

We stress that the symplectic structure of gravitational Cauchy
data is given here by the two first terms on the right-hand side
of (\ref{17all}). Neglecting the second (surface) integral, the
symplectic form would not be gauge invariant with respect to
spacetime diffeomorphisms (see \cite{K1}). The sum of these two
terms plays, therefore, role of the integral over the string
interval $[a,b]$ in formula (\ref{finite}). Most authors analyzing
these problems take only the first (volume) integral as the
symplectic form, which makes the entire approach gauge-dependent.

The last integral (\ref{form-zerowa1}) is responsible for the
control of five components of the  boundary data: the two-metric
$g_{AB}$ on $\partial V$ and the ``curvature'' ${\cal W}_A$.
Assuming the null-like character of $S$ we already control the
sixth parameter: $g^{33}\equiv 0$ (see formula (\ref{gu})). This
corresponds to the general observation (cf.~\cite{K1}) that we
must always control four gauge parameters of the boundary $S$,
together with boundary data of the two ``true degrees of freedom''
of the gravitational field.

\section{Dynamics of gravitational field outside of a null surface}

Consider now dynamics of the gravitational field outside of a wave
front $S^-$. We first add an external, timelike (non-degenerate)
boundary $S^+$ and the situation is illustrated by the following
figure:
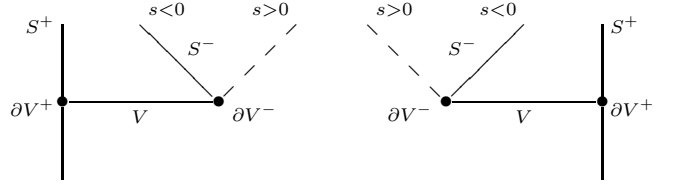
\[%
\xymatrix@M=0pt@R=0.95cm@C=0.95cm{ \ar@{-}[d]_<{S^+} &
\ar@{-}[dr]^<{s<0}^{S^-}& & \ar@{--}[dl]^>{\partial V^-
}_<{s>0}&\ar@{--}[dr]_>{\partial V^- }^<{s>0}&&
\ar@{-}[dl]_<{s<0}_{S^-} &\ar@{-}[d]^<{S^+}\\
{\bullet}\ar@{-}[r]_>V& \ar@{-}[r]&{\bullet}&&&{\bullet}\ar@{-}[r]
& \ar@{-}[r]_<V&{\bullet}\\ \ar@{-}[u]^>{\partial
V^+}&&&&&&&\ar@{-}[u]_>{\partial V^+} }\] where $\partial V^+=
V\cap S^+$, and $\partial V^-=V\cap S^-$. Because $\partial V^- $
enters with negative orientation, we have: $\int_{\partial
V}=\int_{\partial V^+}-\int_{\partial V^- }$. Integrating again
Einstein equations written in the form (\ref{gen-Lang-grav}), over
$V$, we use techniques derived in \cite{pieszy} and \cite{K1} to
handle surface integrals over the timelike surface $S^+$. To
handle surface integrals over $S^-$ we use our formula
(\ref{form-zerowa1a}). This way we obtain:
\begin{align}\label{form-zerowa2}
-& \delta {\cal H}  = -\delta {\cal H^+}-\delta {\cal H^-}
\nonumber\\
&\hspace{-0.5cm}= \frac 1{16 \pi} \int_V \left( {\dot P}^{kl}
\delta g_{kl} - {\dot
g}_{kl} \delta P^{kl} \right) \nonumber \\
 &\hspace{-0.5cm} + \frac{1}{8\pi}\int_{\partial V^+
}\left( \dot{\lambda}\delta\alpha
  -\dot{\alpha}\delta\lambda\right)
+ \frac s{8 \pi} \int_{\partial V^- } ( {\dot \lambda} \delta a  -
{\dot  a} \delta \lambda )\nonumber\\
& \hspace{-0.5cm}-\frac{1}{16\pi}\int_{\partial V^+ }
\mathcal{Q}^{ab}\delta g_{ab}
 \nonumber \\
 &\hspace{-0.5cm}+  \frac s{16 \pi} \int_{\partial V^- } \left( \lambda
{l}^{AB}\delta g_{AB}
 -2\left(w_0\delta\Lambda^0-\Lambda^A\delta w_A
\right)\right),
\end{align}
where $\alpha$ is the ``hyperbolic angle'' between $V$ and $S^+$,
whereas $\mathcal{Q}^{ab}$ is the external curvature of $S^+$
written in the ADM form (cf. \cite{K1}). The contribution ${\cal
H^+}$ to the total Hamiltonian from the external boundary is
written here in the form of a ``free energy'' proposed in
\cite{K1}:
\begin{equation}
{\cal H^+}  = -\frac{1}{8\pi}\int_{\partial V^+} {\mathcal{Q}^0}_0
- E_0 ,
\end{equation}
where the additive gauge $E_0$ is chosen in such a way that the
entire quantity vanishes if $\partial V^+$ is a round sphere in a
flat space. The internal contribution ${\cal H^-}$ to the energy
is given by formula (\ref{H-grav-volume1}) with $\partial V$
replaced by $\partial V^-$. It was proved in \cite{K1} that
shifting the external boundary to space infinity: $\partial V^+
\rightarrow \infty$, the external energy ${\cal H}^+$ gives the
ADM mass, which we denote by $\cal M$, whereas the remaining
surface integrals over $\partial V^+$ vanish. Using this procedure
we obtain the following generating formula for the field dynamics
outside of an arbitrary wave front $S^-$ in an asymptotically flat
spacetime:
\begin{align}\label{form-zerowa21}
 -&\delta {\cal M} -\delta{\cal H^-} = \frac 1{16 \pi} \int_V
\left( {\dot P}^{kl}  \delta g_{kl} - {\dot g}_{kl} \delta P^{kl}
\right)\nonumber\\& + \frac s{8 \pi} \int_{\partial V^- } ( {\dot
\lambda} \delta a - {\dot  a} \delta \lambda )\nonumber \\ & +
\frac s{16 \pi} \int_{\partial V^- } \left( \lambda {l}^{AB}\delta
g_{AB}
 +2\left(\kappa\delta\lambda - n^A\delta {\cal
W}_A \right)\right).
\end{align}

\section{Black hole thermodynamics}\label{czdz}

In this Section we apply the above result to the situation, when
the wave front $S^-$ is a non-expanding horizon, i.e., $l=0$ (see
\cite{jjrw}). In this case the ``internal energy''  ${\cal H^-}$
given by formula \eqref{H-grav-volume1} vanishes. Moreover,
Einstein equations imply ${l}^{AB}=0$ and definition
\eqref{wu-def} of $w_a$ reduces to: $\nabla_a K = -w_a K$ (see
\cite{JKC}). Hence, we obtain the following generating formula for
the black hole dynamics:
\begin{align}
 - \delta {\cal M} & =  \frac 1{16 \pi} \int_V \left( {\dot P}^{kl}
 \delta g_{kl} - {\dot g}_{kl} \delta P^{kl} \right)\nonumber\\
 &  + \frac s{8
 \pi} \int_{\partial V^-} ( {\dot \lambda} \delta a  - {\dot  a}
 \delta \lambda )
 \nonumber \\
 &  +  \frac s{8\pi} \int_{\partial V^-}
 \left(\kappa\delta\lambda - n^A\delta
  {\cal W}_A \right), \label{form-zerowa3}
\end{align}
where $s=1$ for a white hole, and $s=-1$ for a black hole.

The so called "black hole thermodynamics" consists in restricting
the above analysis only to stationary situations. By stationarity
we understand the existence of a timelike symmetry (Killing)
vector field outside of the horizon. If such a field exists,  we
may always choose a coordinate system such that the Killing field
becomes $\tfrac{\partial}{\partial x^0}$ and all the time
derivatives (dots) vanish. Hence, formula (\ref{form-zerowa3})
reduces to:
\begin{equation}\label{form-zerowa11}
 \delta {\cal M}  =
   - \frac s{8 \pi}
  \int_{\partial V^-} \left(\kappa\delta\lambda - n^A\delta
  {\cal W}_A \right).
\end{equation}
Moreover, we assume that $\tfrac{\partial}{\partial x^0}$ is
tangent to $S$. If this was not the case, we would have had a
one-parameter family of horizons. Such phenomenon corresponds to
the Kundt's class of metrics (see e.g.,~\cite{TPJLJJ}). The known
metrics of this class are not asymptotically flat. However, we do
not know whether or not this is a universal property and we
exclude such a pathology by the above assumption.

We have shown in \cite{JKC} that there is a canonical affine
fibration $\pi : S \rightarrow B$ over a base manifold $B$, whose
topology is assumed to be that of a sphere $S^2$. The affine
structure of the fibers is implied by the fact that they are null
geodesic lines in $M$. Identity: $-2{l}_{ab}={\pounds}_K g_{ab}=0$
implies that the metric $g$ on $S$ may be projected onto the base
manifold $B$, which acquires a Riemannian two-metric tensor
$h_{AB}$. The degenerate metric $g_{ab}$ on the manifold $S$ is
simply the pull back of $h_{AB}$ from $B$ to $S$: $g = \pi^* h .$

The quantity $w_a$  is not an intrinsic property of the surface
itself, but depends upon a choice of the null field  $K$ on $S$.
Indeed, if ${\tilde K} = \exp (-\gamma)K$ then $\tilde{w}_a= w_a+
\partial_a \gamma$. In particular, there are on $S$ vector fields
$K$ such that $K^a \nabla_a K = 0$ and, consequently, $\kappa=0$.
They correspond to the affine parametrization of the fibers of
$\pi : S \rightarrow B$.

In case of a black hole, there is a privileged field $K$,
compatible with the timelike symmetry of the solution, which is
normalized to unity at infinity. This way the quantities $\kappa$
and $w_A$ in formula (\ref{form-zerowa11}) become uniquely
defined.

We have, therefore, two symmetry fields of the metric $g_{ab}$ on
$S$: $\partial_0$ and $K$. Due to normalization chosen above, we
have $<\partial_0 - K , dx^0 > = 0$. Hence, the field
$\vec{n}:=\partial_0 - K = n^A \partial_A$ is purely spacelike and
projects on $B$. Moreover, it is a symmetry field of the
Riemannian two-metric $h_{AB}$.

Because the conformal structure of $h_{AB}$ is always isomorphic
to the conformal structure of the unit sphere  $S^2$, we are free
to choose a coordinate system in which  $h_{AB} = f\breve{h}_{AB}$
(and $\breve{h}_{AB}$ denotes the standard unit two-sphere
metric). The field $\vec{n}$ is, therefore, the symmetry field of
this conformal structure. Consequently, $\vec{n}$ belongs to the
six-dimensional space of conformal fields on the two-sphere. Using
remaining gauge freedom, we may choose angular coordinates
$(x^A)=(\theta,\phi)$ in such a way that $\vec{n}$ becomes a
rotation field on the two-sphere. This means (cf.
\cite{lewandowski} or Appendix \ref{omega-proof}) that there
exists a coordinate system in which the following holds:
\begin{equation}\label{Omega1a}
 \vec{n} = -\Omega^k \epsilon_{klm}  {y^l}\partial^m .
\end{equation}
Here $\Omega^k$ are components of a three-dimensional vector
called  angular velocity of the black hole, and  $y^k$ are
functions on $S^2$ created by restricting Cartesian coordinates on
$\mathbb{R}^3$ to a unit two-sphere. We can also set
$z$-coordinate axis parallelly to angular velocity vector field.
After a suitable rotation we have: $(\Omega^k)=( 0,0,\Omega)$,
$z=y^3=\cos\theta$, and:
\begin{equation}\label{Omega1}
  \vec{n}= - \Omega\frac{\partial}{\partial\varphi}.
\end{equation}
Inserting this into (\ref{form-zerowa11}) we obtain
\begin{equation}\label{calka}
  -\frac1{8\pi}\int_{\partial V^-} n^A\delta
  {\cal W}_A =  \Omega \delta J ,
\end{equation}
where
\begin{eqnarray}\label{moment-pedu}
  J \equiv J_z: =
  \frac1{8\pi} \int_{\partial V^-} {\cal W}_\varphi
\end{eqnarray}
is the $z$-component of the black hole angular momentum.

Up to now we have used only the symmetry of conformal structure
carried by  $h_{AB}$. The symmetry of the metric itself implies
that the conformal factor $f$ is constant along  the field
$\vec{n}$. This follows from the observation that the trace of the
Killing equation implies vanishing of divergence of the field
$\vec{n}$:
\begin{equation}\label{divergencja}
  0 = \partial_A (\sqrt{\det h_{CD}}\  n^A)  =
   n^A\sqrt{\det \breve{h}_{CD}}\ \partial_A f
   ,
\end{equation}
where the fact that $\vec{n}$ is the symmetry field of the metric
$\breve{h}$ has been used. Formula (\ref{Omega1}) implies that
$\partial_\varphi f =0$ and the conformal factor $f$ must be a
function of the variable $\theta$ only.

It turns out that also its canonical conjugate $\kappa$ may be
gauged in such a way that it is constant along the field $\vec{n}$
(see Appendix \ref{kappa-proof} for a proof).\footnote{In case
$\Omega=0$, quantities $\kappa$ and $f$ are arbitrary functions on
$S^2$.}

This result was obtained locally, or rather   {\em quasi}-locally
-- i.e., from the analysis of the field on the horizon itself.
However, the {\em global} theorems on the existence of stationary
solutions possessing a horizon, imply the so called zeroth law of
thermodynamics of black holes (see \cite{BH}), according to which
{\em the surface gravity $\kappa$ must be constant along the
horizon}. But $A:=\int_{S^2} \lambda$ is the area of the horizon
$S$. Taking this into account and using (\ref{calka}), we derive
from (\ref{form-zerowa11}) the ``first law of black holes
thermodynamics'':
\begin{equation}\label{form-zerowa22}
 -s\delta {\cal M}  =
   \frac 1{8 \pi}{\kappa}\delta A+\Omega\delta J .
\end{equation}
Contrary to the theory proposed by Wald and Iyer in \cite{Wald},
the first law (\ref{form-zerowa22}) is, in our approach, a simple
consequence of the complete Hamiltonian formula
(\ref{form-zerowa3}), restricted to the stationary case. As
illustrated by an example of the string dynamics, where formula
(\ref{virt-work}) for virtual work was a consequence of the
Hamiltonian formula (\ref{finite}), a similar ``thermodynamics of
boundary data'' may be expected in any Hamiltonian field theory
(see e.g.,~\cite{K1} for the corresponding analysis of the Maxwell
electrodynamics). Also a ``Penrose-like'' inequality (analogous to
(\ref{ineq}) in the string theory) is satisfied as soon as the
Hamiltonian is convex. We hope very much that the gravitational
Penrose inequality can be proved along these lines. Preliminary
results in this direction, based on the analysis of the field
Hamiltonian in linearized gravity (see \cite{JJGRG31}), are
promising.

\section{Acknowledgements}
This research was supported in part by the Polish Research Council
grant KBN 2 P03B 073 24 and by the Erwin Schr\"odinger Institute.

\appendix

\section{Notation and preliminary formulae}\label{framework}

A considerable simplification of the proofs is obtained if we use
in a neighbourhood of $S=\{x^3 = {\rm const}\}$ a special
coordinate system introduced in \cite{JKC}, which reduces the
metric to the following special form:
\begin{equation}\label{gd}
{g}_{\mu\nu} = \left[ \begin{array}{ccccc} n^A n_A & \vline & n_A
& \vline & sM+m^A n_A \\
 & \vline &  & \vline & \\
\hline & \vline &  & \vline &  \\
 n_A & \vline & g_{AB} & \vline & m_A \\
 & \vline &  & \vline & \\
 \hline & \vline &  & \vline &  \\
 sM+m^A n_A & \vline & m_A & \vline &
 \left( \frac{M}{N}\right)^2+m^A m_A \\
 \end{array} \right] .
\end{equation}
Consequently, we have
\begin{widetext}
\begin{equation}\label{gu}
{g}^{\mu\nu} = \left[ \begin{array}{ccccc} - \left( \frac 1N
\right)^2 & \vline & \frac {n^A}{N^2} - s\frac {m^A}{M} & \vline &
\frac {s}{M} \\
 & \vline &  & \vline & \\
\hline & \vline &  & \vline &  \\
 \frac {n^A}{N^2} - s\frac {m^A}{M} & \vline &
 {\tilde{\tilde g}}^{AB} - \frac {n^A n^B}{N^2} + s\frac {n^A m^B + m^A
 n^B}{M} & \vline & - s\frac {n^A}{M} \\
 & \vline &  & \vline & \\
 \hline & \vline &  & \vline &  \\
 \frac s{M} & \vline & - s\frac {n^A}{M} & \vline & 0 \\
 \end{array} \right] ,
\end{equation}
\end{widetext}
where $M > 0$, $s:=\mathop {\rm sgn}\nolimits g^{03}=\pm 1$,
$g_{AB}$ is the induced two-metric on surfaces $\{x^0={\rm
const},\; x^3={\rm const} \}$ and ${\tilde{\tilde g}}^{AB}$ is its
inverse (contravariant) metric. Both ${\tilde{\tilde g}}^{AB}$ and
$g_{AB}$ are used to rise and lower indices $A,B = 1,2$ of the
two-vectors $n^A$ and $m^A$.

We denote by $\lambda$ the two-dimensional volume form on each
two-surface $\{x^0={\rm const},\; x^3={\rm const} \}$:
\begin{equation}\label{lambda}
  \lambda:=\sqrt{\det g_{AB}}\, .
\end{equation}
For any degeneracy field $K$ of $g_{ab}$ the following object:
\[
v_{K} := \frac {\lambda}{K(x^0)}
\]
is a scalar density on $S$. The vector density
\begin{equation}\label{Lambda}
\Lambda = v_K \, K = \lambda (\partial_0-n^A\partial_A)\, ,
\end{equation}
is well defined (i.e., ~coordinate-independent) and, obviously,
does not depend upon any choice of the field $K$. Hence, it is an
intrinsic property of $S$.

The external geometry of $S$ is described  in terms of the
following tensor density:
\begin{equation}\label{Q-fund}
{Q^a}_b (K) := -s \left\{ v_K \left( \nabla_b K^a - \delta_b^a
\nabla_c K^c \right) + \delta_b^a \partial_c \Lambda^c \right\} \
\end{equation}
which is  fully analyzed in our previous paper \cite{JKC1}.

In our calculations we shall use also quantities which {\em are
not} geometric objects (are coordinate dependent). All of them
drop out in the final result, where only well defined geometric
objects remain. More precisely, we consider the following
combination of the connection coefficients:
\begin{align} \label{tQ}
{\widetilde  Q}^{\mu\nu} := \sqrt{|g|}& \left( g^{\mu \alpha}
g^{\nu \beta} - \frac 12 g^{\mu\nu} g^{ \alpha \beta}
\right)\nonumber\\
&\times \left(\Gamma^3_{\alpha\beta}-\delta^3_{\alpha}
\Gamma^{\lambda}_{\beta\lambda}\right) ,
\end{align}
and the two-dimensional inverse metric ${\tilde{\tilde g}}^{AB}$
rewritten in a ``three-dimensional notation'', where we put
${\tilde{\tilde g}}^{0a} := 0$. The resulting matrix
${\tilde{\tilde g}}^{ac}$ does not define any tensor on $S$ and
satisfies the obvious identity:
\[
 {\tilde{\tilde g}}^{ac} g_{cb}=\delta^a{_b}-K^a\delta^0{_b} \, .
\]
Hence, the $(0,1,2)$-block of the contravariant metric (\ref{gu})
may be rewritten as follows:
\begin{equation}\label{g2i}
  g^{ab} = {\tilde{\tilde g}}^{ab} -\frac1{N^2} K^aK^b -
 \frac{s}M ( m^a K^b + m^b K^a ) ,
\end{equation}
where $m^a:= {\tilde{\tilde g}}^{aB}m_B$, so that $m^0:=0$, and
\[ g^{3\mu} = \frac sM K^\mu . \]
Using the above definition we may write that
\begin{align}\label{Pli}
s {\tilde{Q}}^a{_b} = &\lambda \left( {\tilde{\tilde
g}}^{ac}l_{cb} -\frac12 \delta^a{_b} l\right) + \Lambda^a w_{b} -
\delta^a{_b} \Lambda^c w_{c}\nonumber\\& + \Lambda^a \chi_b
-\delta^a{_b} \Lambda^c \chi_c \, ,
\end{align}
where $l_{ab},w_{a}$ are defined by (\ref{el-def}) and
(\ref{wu-def}) correspondingly,  $\displaystyle \chi_c := \frac12
\partial_c \ln \left(\frac M\lambda \right)$. From (\ref{Q-fund})
we also have the following
\begin{align}\label{Qli}
 s&Q^a{_b}  =  \lambda\delta^a{_b}\nabla_c K^c -\lambda \nabla_b K^a -
 \delta^a{_b} \partial_c \Lambda^c \nonumber\\
 &= -\lambda\delta^a{_b}(w_cK^c+l) + \lambda (w_bK^a+ {\tilde{\tilde g}}^{ac}
 l_{cb}) + \delta^a{_b} \lambda l \nonumber\\
 &= \lambda  {\tilde{\tilde g}}^{ac} l_{cb}  +
    \Lambda^a w_b -\delta^a{_b} \Lambda^c w_c.
\end{align}
Equation (\ref{Qli}) expresses the $(2+1)$-decomposition of the
three-dimensional density $Q^a{_b}$. As a consequence of
\eqref{Pli} and \eqref{Qli} we obtain the following identity:
\begin{equation}\label{PQ}
 s {\tilde{Q}}^a{_b} = s Q^a{_b} -\frac12 \lambda l \delta^a{_b}+ \Lambda^a
\chi_b -\delta^a{_b} \Lambda^c \chi_c .
\end{equation}
The detailed proof of these formulae is contained in paper
\cite{JKC1}, where we derive also the following equality:
\begin{align}
s{\widetilde Q}^{\alpha\beta}& g_{\alpha\beta ,a}  =
  \lambda
 ( g^{be}g^{cd}l_{ed} -\frac12 l g^{bc})
 g_{bc ,a}\nonumber
 \\
& + (\Lambda^b g^{cd} + \Lambda^c g^{bd}
 -\Lambda^d g^{cb})A^3_{3d} g_{bc ,a} \nonumber\\
 &  + 2 s{\widetilde Q}^3{_3} \left( \partial_a \ln M +\frac{s}M m_B
n^B_{,a}\right) . \label{Pgfull}
\end{align}
Replacing the partial derivative $\partial_a$ by the variation
operator $\delta$, we get an analogous formula
\begin{align*} \nonumber
s{\widetilde Q}^{\alpha\beta}&\delta g_{\alpha\beta }  =
 \lambda
 ( g^{be}g^{cd}l_{ed} -\frac12 l g^{bc})\delta g_{bc
 }\nonumber\\
& + (\Lambda^b g^{cd} + \Lambda^c g^{bd}
 -\Lambda^d g^{cb})A^3_{3d}  \delta g_{bc } \\
 &  + 2 s{\widetilde Q}^3{_3} \left( \frac{1}{M}\delta  M +\frac{s}M m_B
\delta n^B\right),
\end{align*}
which may be further simplified to
\begin{align}\nonumber
s\tilde{Q}^{\mu\nu}\delta g_{\mu\nu}   =&sQ^{ab}\delta g_{ab}-\frac{1}%
{2}\lambda l \tilde{\tilde{g}}^{ab}\delta g_{ab}-\lambda
l\delta\log M\\ \label{tq-war}
&  +\chi_{d}\left(  \Lambda^{a}\tilde{\tilde{g}}^{bd}+\Lambda^{b}%
\tilde{\tilde{g}}^{ad}-\Lambda^{d}\tilde{\tilde{g}}^{ab}\right)
\delta g_{ab}\, .
\end{align}
Moreover, we need the following identities from \cite{JKC1}:
\begin{equation} \label{tq-war1}
\tilde{Q}^{\mu\nu}g_{\mu\nu}=-s\lambda\left(  l+2K^{d}(w_{d}+\chi
_{d})\right),
\end{equation}
and
\begin{equation}\label{ldotjj}
\lambda l = -\dot{\lambda} + \partial_{A}(n^{A}\lambda) \, .
\end{equation}

\section{Proofs of the generating formulae for
dynamics of gravitational field}\label{inside}

Dynamics of gravitational field is derived from the principle of
the least action  $\delta {\cal A}= 0$, where the action of
gravitational field  is defined as the integral of Hilbert
Lagrangian:
\begin{equation}
  L =   \frac 1{16 \pi} \sqrt{|g|} \ R
 .
\label{Hilbert5}
\end{equation}
The method proposed by one of us in papers \cite{pieszy,K1} leads
to Einstein equations written in the following form:
\begin{equation}\label{divergence-A5}
  \delta L =
   \partial_\kappa \left( \pi^{\mu\nu}\delta
   A^\kappa_{\mu\nu} \right)
   ,
\end{equation}
where
\begin{align*}{\pi}^{\mu\nu} :&= \frac 1{16
\pi} \sqrt{|g|} \ g^{\mu\nu}, \\A^{\lambda}_{\mu\nu} :&=
{\Gamma}^{\lambda}_{\mu\nu} - {\delta}^{\lambda}_{(\mu}
{\Gamma}^{\kappa}_{\nu ) \kappa}.
\end{align*}
As soon as we choose a (3+1)-decomposition of the spacetime $M$,
our field theory will be converted into a Hamiltonian system, with
the space of Cauchy data on each of the three-dimensional surfaces
playing role of an infinite-dimensional phase space. Let us choose
coordinate system adapted to this (3+1)-decomposition. This means
that the time variable $t = x^0$ is constant on three-dimensional
surfaces of this foliation. We assume that these surfaces are
spacelike. To obtain Hamiltonian formulation of our theory we
shall simply integrate equation (\ref{divergence-A5}) over such a
Cauchy surface ${\cal C}_t \subset M$ and then perform Legendre
transformation between time derivatives and corresponding momenta.

We consider the case of an asymptotically flat spacetime and
assume that also leaves ${\cal C}_t$ of our (3+1)-decomposition
are asymptotically flat at infinity. To keep control over
two-dimensional surface integrals at spatial infinity, we first
describe dynamics of our ``matter + gravity'' system within a
finite volume $V$. Integration of (\ref{divergence-A5}) over $V$
yields:
\begin{align}\label{generate-205}
  \delta\int_V  L &=\int_V \partial_\kappa \left(
  \pi^{\mu\nu}\delta
   A^\kappa_{\mu\nu} \right)
   =\nonumber\\
   &\int_V\left( \pi^{\mu\nu}\delta
   A^0_{\mu\nu} \right)^\cdot + \int_{\partial V}
   \pi^{\mu\nu}\delta
   A^\perp_{\mu\nu}
 ,
\end{align}
where   ``dot'' denotes time derivative (the two-dimensional
divergence $\partial_B \left( \pi^{\mu\nu}\delta A^B_{\mu\nu}
\right)$ vanishes when integrated over $\partial V$).

In \cite{pieszy} the Legendre transformation between time
derivatives and corresponding momenta was performed in case of a
non-degenerate (one-timelike, two-spacelike) surface $S$. Here we
generalize this method to the case of a wave front. The first step
in this construction consists in observation that, due to
metricity of the connection $\Gamma$, the following identity
holds:
\begin{equation}\label{redukcja-0}
    \pi^{\mu\nu}\delta A^0_{\mu\nu}=
    -\frac{1}{16\pi}g_{kl}\delta P^{kl} +
    \partial_k\left(\pi^{00}\delta\left(
    \frac{\pi^{0k}}{\pi^{00}}\right)\right) ,
\end{equation}
where $P^{kl}$ denotes the external curvature of $\Sigma$ written
in the ADM~form. A simple proof of this formula is also contained
in paper \cite{JKC1}.

On the other hand, direct calculations of the variation of  the
quantity ${\tilde{Q}}^{\mu\nu}$ given by  (\ref{tQ}) lead to the
following reduction of the boundary term $\pi^{\mu\nu}\delta
A^\perp_{\mu\nu}=\pi^{\mu\nu}\delta A^3_{\mu\nu}$:
\begin{equation}\label{redukcja-3}
    \pi^{\mu\nu}\delta A^3_{\mu\nu}=
    -\frac{1}{16\pi}g_{\mu\nu}\delta {\tilde{Q}}^{\mu\nu}  .
\end{equation}

Skipping the two-dimensional divergencies which vanish after
integration and using (\ref{redukcja-0}) and (\ref{redukcja-3}),
we may rewrite the right-hand side of (\ref{generate-205}) in the
following way:
\begin{align}\nonumber
   \int_V&\left( \pi^{\mu\nu}\delta
   A^0_{\mu\nu} \right)^\cdot + \int_{\partial V}
   \pi^{\mu\nu}\delta
   A^\perp_{\mu\nu} \nonumber\\   = &
   -\frac{1}{16\pi}\int_V\left( g_{kl}\delta P^{kl} \right)^\cdot
   +  \int_{\partial  V}\left(\pi^{00}\delta\left(
    \frac{\pi^{03}}{\pi^{00}}\right)\right)^\cdot\nonumber\\
    &
    -\frac{1}{16\pi} \int_{\partial V} g_{\mu\nu}\delta {\tilde{Q}}^{\mu\nu} .
\end{align}
Now, we perform the Legendre transformation both in the volume:
\be \left(  g_{kl}\delta P^{kl}\right)  ^{\cdot}=\left(
\dot{g}_{kl}\delta P^{kl}-\dot{P}^{kl}\delta g_{kl}\right)
+\delta\left(  g_{kl}\dot{P} ^{kl}\right) \ee
 and on the boundary:
\begin{align}
& \left(\pi^{00}\delta\left(
    \frac{\pi^{03}}{\pi^{00}}\right)\right)^\cdot=
 \left(\pi^{00}\right)^\cdot\delta\left(
    \frac{\pi^{03}}{\pi^{00}}\right)\nonumber\\&-\left(
    \frac{\pi^{03}}{\pi^{00}}\right)^\cdot\delta\pi^{00}+
    \delta\left(\pi^{00}\left(
    \frac{\pi^{03}}{\pi^{00}}\right)^\cdot \ \right).
\end{align}
In paper \cite{JKC1} the following formula has been proved:
\begin{align}
\label{pe-alpha5} &- \int_{V}\left( g_{kl}\dot P^{kl}\right)
+16\pi\int_{\partial V}\pi^{00}\left(
    \frac{\pi^{03}}{\pi^{00}}\right)^\cdot\nonumber\\
    &=\left\{  \partial_{k}\left(  \sqrt{|g|}\left(  g^{k\mu}%
\Gamma_{0\mu}^{0}-g^{0\mu}\Gamma_{0\mu}^{k}\right)  \right)
+2\sqrt {|g|}{R^{0}}_{0}\right\}.
\end{align}
Hence, generating formula (\ref{generate-205}) takes the form:
\begin{align}
\delta\int_{V} &  L=\frac{1}{16\pi}\int_{V}\left(
\dot{P}^{kl}\delta g_{kl}-\dot{g}_{kl}\delta
P^{kl}\right)\nonumber\\& -\frac{1}{16\pi} \int_{\partial V}
g_{\mu\nu}\delta
{\tilde{Q}}^{\mu\nu} \nonumber\\
&  +\int_{\partial V}\left(  \dot{\pi}^{00}\delta\left(  \frac{\pi^{03}}%
{\pi^{00}}\right)  -\left(  \frac{\pi^{03}}{\pi^{00}}\right)  ^{\dot{}}%
\delta\pi^{00}\right)\nonumber\\
&+\frac{1}{16\pi}\delta\int_{\partial V}\sqrt {|g|}\left(
g^{3\mu}\Gamma_{0\mu}^{0}-g^{0\mu}\Gamma_{0\mu}^{3}\right)\nonumber\\
&+\frac{1}{8\pi}\delta\int_{V}\sqrt{|g|}{R^{0}}_{0}.
\end{align}
Using the form of the metric  (\ref{gd}) and  (\ref{gu}) we
express $\pi^{\mu\nu}$ in terms of the metric. Denoting
\begin{equation}
a:=\log N
\end{equation}
we obtain
\begin{align} \nonumber
\int_{\partial V}&   \left(  \dot{\pi}^{00}\delta\left(  \frac{\pi^{03}}%
{\pi^{00}}\right)  -\left(  \frac{\pi^{03}}{\pi^{00}}\right)  ^{\cdot}%
\delta\pi^{00}\right)  \\   =&\frac{s}{16\pi}\int_{\partial
V}\left( 2\dot{\lambda}\delta a -2\dot{
a}\delta\lambda\right)\nonumber\\
&-\frac{s}{16\pi}\int_{\partial
V}\left(\dot{\lambda}\delta{\log M}-(\log M)^{\cdot}%
\delta\lambda\right) .
\end{align}
Similarly, we rearrange the boundary term:
\[
g_{\mu\nu}\delta\tilde{Q}^{\mu\nu} = \delta\left(
g_{\mu\nu}\tilde{Q}^{\mu\nu }\right)  -\tilde{Q}^{\mu\nu}\delta
g_{\mu\nu} .
\]
For this purpose we use the formula (\ref{tq-war}) and
(\ref{tq-war1}). Therefore, our generating formula takes the
following form:
\begin{align}\label{gen-form05}
&\frac{1}{16\pi}  \delta\int_{V}\sqrt{g}\left(
R-2R^{0}{_{0}}\right)
=-\frac{1}{8\pi}\delta\int_{V}\mathcal{G}^{0}{_{0}}\nonumber\\&=
\frac{1}{16\pi}\int _{V}\left(  \dot{P}^{kl}\delta g_{kl}-
\dot{g}_{kl}\delta P^{kl}\right) \nonumber \\ &
+\frac{s}{8\pi}\int_{\partial V}\left( \dot{\lambda}\delta a
-\dot{ a}\delta\lambda\right)  +\frac{1}{16\pi}\int_{\partial V}%
Q^{ab}(K)\delta g_{ab}\nonumber\\ & +\frac{s}{16\pi}\int_{\partial
V}
\left[-\frac{1}{2}\lambda\tilde{\tilde{g}}%
^{ab}\delta g_{ab}-\lambda l\delta\log M-\dot{\lambda}\delta\log
M\right.\nonumber\\ &\quad \left.+(\log M)^{\cdot}\delta\lambda
+\chi_{d}\left(
\Lambda^{a}\tilde{\tilde{g}}^{bd}+\Lambda^{b}\tilde
{\tilde{g}}^{ad}-\Lambda^{d}\tilde{\tilde{g}}^{ab}\right) \delta
g_{ab}\vphantom{\frac12}\right]\nonumber\\ &
+\frac{1}{16\pi}s\delta\int_{\partial V}\lambda
\left(  2K^{d}(w_{d}%
+\chi_{d})+l\right) \nonumber\\
&+\frac{1}{16\pi}\delta\int_{\partial V}\sqrt{|g|}\left(
g^{3\mu}\Gamma_{0\mu}^{0}-g^{0\mu}\Gamma_{0\mu}^{3}\right).
\end{align}
Now, we simplify the following boundary term (in square brackets
in the above formula):
\begin{align}
-\frac{1}{2}\lambda &\tilde{\tilde{g}}%
^{ab}\delta g_{ab}-\lambda l\delta\log M-\dot{\lambda}\delta\log
M+(\log M)^{\cdot}\delta\lambda\nonumber\\
 &+\chi_{d}\left(
\Lambda^{a}\tilde{\tilde{g}}^{bd}+\Lambda^{b}\tilde
{\tilde{g}}^{ad}-\Lambda^{d}\tilde{\tilde{g}}^{ab}\right)  \delta
g_{ab}. \label{wyraz1}
\end{align}
Using identity (\ref{ldotjj}) (see e.g., \cite{JKC}),
$\Lambda^{b}\delta g_{bc}=-g_{bc}\delta \Lambda^{b}$, and skipping
two-dimensional divergencies, we obtain that the above expression
takes the following form:
\begin{align}
&\lambda(\partial_A\log M)\delta n^A
-\partial_A(n^A\lambda)\delta\log M\nonumber\\&+
n^A\partial_A(\log M)\delta \lambda+ (\partial_A n^A)\delta
\lambda-\lambda(\partial_A \log\lambda)\delta n^A .
\end{align}
Further simplifications can be made after integration over
$\partial V$:
\begin{align}
\int_{\partial V}&\left( \partial_A(\log M)\delta
\Lambda^A-(\partial_A \Lambda^A)\delta \log M\right)\nonumber\\
&=-\delta\int_{\partial V}\left( (\partial_A\Lambda^A)\log
M\right),\\ \int_{\partial V}&\left( (\partial_A
n^A)\delta\lambda-(\partial_A\lambda)\delta n^A
\right)\nonumber\\&= \int_{\partial V}\left(\partial_A
n^A)\delta\lambda+\lambda\delta(\partial_A n^A)\right)\nonumber\\
& =\delta\int_{\partial V}\lambda(\partial_A n^A) .
\end{align}
Finally, expression (\ref{wyraz1}) takes  the following form ({\em
modulo} two-dimensional divergencies):
\[
\delta\left( (\partial_A\Lambda^A)\log M + \lambda\partial_A
n^A\right),
\]
and  the formula (\ref{gen-form05}) reads as
\begin{align}\label{gen-form056}
-\frac{1}{8\pi} &  \delta\int_{V}{\cal G}^0{_0}=
\frac{1}{16\pi}\int _{V}\left(  \dot{P}^{kl}\delta g_{kl}-
\dot{g}_{kl}\delta P^{kl}\right) \nonumber \\ &
+\frac{s}{8\pi}\int_{\partial V}\left( \dot{\lambda}\delta a
-\dot{ a}\delta\lambda\right)  +\frac{1}{16\pi}\int_{\partial V}%
Q^{ab}(K)\delta g_{ab}\nonumber\\ &+\frac{s}{16\pi}\delta
\int_{\partial V}\left( (\partial_A\Lambda^A)\log M +
\lambda\partial_A n^A\right)\nonumber\\
&  +\frac{s}{16\pi}\delta\int_{\partial V}\lambda\left(  2K^{d}(w_{d}%
+\chi_{d})+l\right)\nonumber\\&
+\frac{1}{16\pi}\delta\int_{\partial V}\sqrt{|g|}\left(
g^{3\mu}\Gamma_{0\mu}^{0}-g^{0\mu}\Gamma_{0\mu}^{3}\right).
\end{align}
Now, we simplify the expression:
\begin{align}\label{wyraz2}
\delta\int_{\partial V}&\left\{s(\partial_A\Lambda^A)\log M +
s\lambda\partial_A n^A\right.\nonumber\\& + s\lambda\left(
2K^{d}(w_{d} +\chi_{d})+l\right)\nonumber\\&\left.
+\sqrt{|g|}\left(
g^{3\mu}\Gamma_{0\mu}^{0}-g^{0\mu}\Gamma_{0\mu}^{3}\right)\right\}.
\end{align}
Using \be w_a=\Gamma^3_{3a} - \frac sM m^c l_{ca} - \frac 1M
  M_{,a}
\ee
 and taking into account that $K^\mu\Gamma^0_{\mu
a}=-w_a$ (\ref{wu-def}), $K_\mu\Gamma^\mu_{ab}=l_{ab}$
(\ref{el-def}) we can pass to the metric derivatives contained in
$\Gamma$'s. This way (\ref{wyraz2}) reduces ({\em modulo}
two-dimensional divergencies) to
\begin{equation}
-2s\int_{\partial V} l\delta\lambda+  2s\delta\int_{\partial
V}\lambda\left(l-n^{A}w_{A}\right).
\end{equation}
Therefore, we obtain the following boundary formula:
\begin{align} \nonumber
-&\frac{1}{8\pi}
\delta\int_{V}\mathcal{G}^{0}{_{0}}=\frac{1}{16\pi}\int _{V}\left(
\dot{P}^{kl}\delta g_{kl}-\dot{g}_{kl}\delta P^{kl}\right)  \\ &
+\frac{s}{8\pi}\int_{\partial V}\left(  \dot{\lambda}\delta a
-\dot{ a}\delta\lambda\right)\nonumber\\&
+\frac{1}{16\pi}\int_{\partial V}\left( Q^{ab}(K)\delta
g_{ab}-2sl\delta\lambda\right) \nonumber\\&+\frac{s}{8\pi}\delta
\int_{\partial V}\lambda(l-n^{A}w_{A}).
\end{align}
Using identities  $ Q^{ab}(K)\delta g_{ab}=2Q^{0}{_{A}}(K)\delta
n^{A}-Q_{AB}(K)\delta\tilde{\tilde{g} }^{AB}$ and $s\lambda
n_{A}w^{A}=Q^{0}_{A}(K) n^{A}$ we get:
\begin{align}\label{form-zerowa} -&\frac{1}{8\pi}
\delta\int_{V}\mathcal{G}^{0}{_{0}}=\frac{1}{16\pi}\int _{V}\left(
\dot{P}^{kl}\delta g_{kl}-\dot{g}_{kl}\delta P^{kl}\right)
\nonumber\\ & +\frac{s}{8\pi}\int_{\partial V}\left(
\dot{\lambda}\delta a -\dot{ a}\delta\lambda\right)\nonumber\\
-\frac{1}{16\pi}&\int_{\partial V}\left(
Q_{AB}(K)\delta\tilde{\tilde{g}}^{AB}+2n^{A}\delta
Q^{0}{_{A}}(K)-2s\lambda\delta l\right).
\end{align}
Using equation (\ref{Qli}) we express $Q$ in terms of independent
objects $l_{ab}$ and $w_a$, and finally obtain:
\begin{align}
- &\delta {\cal H}  =  \frac 1{16 \pi} \int_V  \left( {\dot
P}^{kl} \delta g_{kl} - {\dot g}_{kl} \delta P^{kl}
\right)\nonumber\\& + \frac s{8 \pi} \int_{\partial V} ( {\dot
\lambda} \delta  a  -
{\dot  a} \delta \lambda ) \nonumber \\
 & +  \frac s{16 \pi}
\int_{\partial V} \left( \lambda {l}^{AB}\delta g_{AB}
 -2\left(w_0\delta\Lambda^0-\Lambda^A\delta w_A
\right)\right),
\end{align}
where
\begin{equation}
{\cal H}  =   \frac 1{8 \pi} \int_V {\cal G}^0_{\ 0} + \frac s{8
\pi}
 \int_{\partial V}  \lambda l
\equiv   \frac s{8 \pi}
 \int_{\partial V}  \lambda l
  .
\end{equation}

\section{Derivation of generating formula from the non-degenerate
case}

Consider a one-parameter family of hypersurfaces
$S_\epsilon:=\{r_\epsilon :=r-s\, \epsilon\, t={\rm const}\}$
parameterized by a real  $\epsilon$, such that for  $\epsilon=0$
we have $S_0= S$ and for each $\epsilon\neq 0$ the induced
three-metric on $S_\epsilon$ is non-degenerate.

Take the external curvature tensor  ${\cal Q}^{ab}$ of
$S_\epsilon$ and the three-dimensional contravariant metric
$\hat{g}^{ab}$, inverse to $g_{ab}$. Define
\begin{align}
{\cal Q} & :=\frac{1}{\sqrt{|\hat{g}^{00}|}}{\cal Q}^{00}, \\
{\cal Q}_A &:={\cal Q}^0{_A},
\\ \stackrel{\perp\ \ \ }{{\cal Q}^{AB}} & := {\cal
Q}_{CD}\tilde{\tilde{g}}^{CA} \tilde{\tilde{g}}^{DB}.
\end{align}
Observe that these are two-dimensional objects defined on
$\partial V$: ${\cal Q}$ is a scalar density, ${\cal Q}_A$ is a
covector density and $\stackrel{\perp\ \ \ }{{\cal Q}^{AB}}$ is a
symmetric tensor density. The following identity (a homogeneous
generating formula for the field dynamics) was proved in \cite{K1}
for a timelike boundary and generalized in \cite{Kasia} for any
non-degenerate (e.g.,~a spacelike) boundary:
\begin{align}\label{homog-form}
0= \int_{V}&\left(  \dot{P}^{kl}\delta g_{kl}-\dot{g}%
_{kl}\delta P^{kl}\right)  +2\int_{\partial V}\left(
\dot{\lambda}\delta\alpha
  -\dot{\alpha}\delta\lambda\right)\nonumber\\
&+\int_{\partial V}(\stackrel{\perp\ \ \ }{{\cal Q}^{AB}}\delta
g_{AB}-2n^A\delta {\cal Q}_A +2n\delta { \cal Q}),
\end{align}
where
\begin{align*}
n:=&\frac{1}{\sqrt{|\hat{g}^{00}|}},\\ n^A:=&\tilde{\tilde
{g}}^{AB}g_{0B}.
\end{align*}
The boundary objects  $\alpha$, ${\cal Q}$, ${\cal Q}_A$ and
$\stackrel{\perp\ \ \ }{{\cal Q}^{AB}}$ diverge under the limit
$S_\epsilon\rightarrow S$ -- or, equivalently, $g^{33}\rightarrow
0$. We are going to rearrange them in such a way that the
resulting objects do behave well under this limit.

For this purpose three-dimensional metric component $\hat{g}^{00}$
defining the lapse function $n$, can be expressed by the following
components of the spacetime metric:
\begin{equation}
\hat{g}^{00}=g^{00}-\frac{(g^{03})^2}{g^{33}}=
\frac{g^{00}g^{33}-(g^{03})^2}{g^{33}}.
\end{equation}
Four-dimensional component  $g^{33}$ of the inverse metric can be
expressed in terms of three-dimensional metric component
$\tilde{g}^{33}$  inverse to metric  $g_{kl}$ induced on $V$:
\begin{equation}
g^{33}=\tilde{g}^{33}+\frac{(g^{03})^2}{g^{00}}.
\end{equation}
Then
\begin{equation}
\hat{g}^{00}=\frac{g^{00}\tilde{g}^{33}}{g^{33}}.
\end{equation}
Restricting formula \eqref{gd} to the hypersurface $\{ x^0 = {\rm
const}\}$ we obtain:
\begin{equation*}
g_{kl}=\left[
\begin{array}{ccc} \displaystyle
g_{AB} & \vline &m_A\\ & \vline & \\ \hline & \vline & \\ m_A &
\vline & \left( \frac{M}{N}\right)^2+m^A m_A
\end{array}
\right] .
\end{equation*}
Consequently, its three-dimensional inverse equals:
\begin{equation*}
\tilde{g}^{kl}=\left(\frac{N}{M}\right)^2\left[
\begin{array}{ccc} 
\left(\left(\frac{M}{N}\right)^2+m^A
m_A\right)\tilde{\tilde{g}}^{AB} & \vline &-m^A\\ & \vline & \\
\hline & \vline & \\ -m^A & \vline & 1
\end{array}
\right] .
\end{equation*}
In this notation $g^{00}=-1/N^2$, hence we obtain that
$\hat{g}^{00}=-{1}/({M^2 g^{33}})$ and the lapse $n$ equals
\begin{equation}
  n=M\sqrt{g^{33}}.
\end{equation}

Using definition of the hyperbolic angle $\alpha = {\rm arsinh}
(q)$, where $\displaystyle q =\frac{g^{30}}{\sqrt{|
g^{00}g^{33}|}}$, we obtain:
\begin{equation}
\delta\alpha=\frac{\delta
q}{\sqrt{1+q^2}}=\frac{1}{\sqrt{1+q^{-2}}}\delta(\log q) .
\end{equation}
Taking into account that
\begin{equation}
\log q=\log g^{03}-\frac12(\log |g^{00}| +\log g^{33}),
\end{equation}
we obtain:
\begin{align}\label{alpha1}
\frac q{\sqrt{1+q^2}}\delta\alpha&=\delta\log N - \delta\log
\left( M \sqrt{g^{33}} \right)\nonumber\\&= \delta{\mathfrak a}-
\delta\log n ,
\end{align}
where
\begin{equation}
{\mathfrak a}:= - \log\sqrt{|g^{00}|} \equiv \log N
\end{equation}
is a regular part of $\alpha$. The second term $\delta\log n$
diverges in the limit, but cancels the divergent part of ${\cal
Q}$, what will be proved in the sequel.

Moreover, we have:
\begin{equation}
n{\cal Q}=n^2 {\cal Q}^{00}=\lambda M
\Gamma^3_{AB}\tilde{\tilde{g}}^{AB}=\lambda l ,
\end{equation}
hence the following formula holds:
\begin{equation}
2n\delta {\cal Q}=2n{\cal Q}\delta\left(\log {\cal
Q}\right)=2\lambda l\delta\left(\log\lambda l-\log n\right).
\end{equation}

Continuing, we have:
\begin{align}
{\cal Q}_A&={\cal Q}^0_{\;\;A}=\tilde{g}^{0b} {\cal
Q}_{bA}\nonumber\\& =\lambda
M\left(g^{0b}-\frac{g^{30}g^{3b}}{g^{33}}\right)
\left(\Gamma^{3}_{bA}-g_{bA}\Gamma^3_{cd}\tilde{g}^{cd}\right).
\end{align}
Using the following formula:
\begin{equation}
-\frac12 (\log g^{33})_{,a}=\Gamma^3_{a3}+\frac{1}{g^{33}}
\Gamma^3_{ab}g^{3b} ,
\end{equation}
we obtain
\begin{equation}
{\cal
Q}^0_{\;\;A}=\lambda\left(\Gamma^3_{3A}-m^B\Gamma^3_{BA}\right)
+\frac12\lambda\left(\log g^{33}\right)_{,A}.
\end{equation}
Now, we define the quantity:
\begin{equation}
{\mathfrak w}_a:=-\frac{g^{3\mu}}{g^{30}}\Gamma^0_{\mu a} .
\end{equation}
Taking into account that
\begin{equation}
\Gamma^3_{3a}-m^B\Gamma^3_{Ba}\equiv {\mathfrak w}_a+(\log
M)_{,a},
\end{equation}
it follows that  ${\cal Q}^0_{\;\;A}$ may be written in the
following form:
 \begin{equation}
{\cal Q}^0_{\;\;A}=\lambda{\mathfrak w}_A+\lambda(\log
M)_{,A}+\frac{\lambda}2 (\log g^{33})_{,A}.
\end{equation}

Similarly, $\stackrel{\perp\ \ \ }{{\cal Q}^{AB}}$ takes the
following form:
\begin{align}
\stackrel{\perp\ \ \ }{{\cal
Q}^{AB}}&=Q_{CD}\tilde{\tilde{g}}^{CA}\tilde{\tilde{g}}^{DB}\nonumber\\&
=\lambda M \tilde{\tilde{g}}^{CA}\tilde{\tilde{g}}^{DB}
\left(\Gamma^3_{CD}-g_{CD}\Gamma^3_{ef} {\tilde{g}}^{ef}\right)\,
,
\end{align}
hence the contraction $\stackrel{\perp\ \ \ }{{\cal
Q}^{AB}}g_{AB}$ gives us
\begin{align*}
\stackrel{\perp\ \ \ }{{\cal Q}^{AB}}g_{AB}&=-\lambda
l-2\lambda{\mathfrak w}_0\nonumber\\&+2\lambda n^A {\mathfrak w}_A
- 2\lambda (\partial_0 -n^A
\partial_A)\log M\\ & - \lambda (\partial_0 -n^A
\partial_A)(\log g^{33}),
\end{align*}
and the traceless part of $\stackrel{\perp\ \ \ }{{\cal Q}^{AB}}$
takes the form:
\begin{equation}
 \stackrel{\perp}{{\cal  Q}}\!^{AB}-\frac12
\stackrel{\perp}{{\cal  Q}}\!{^{CD}}g_{CD}\tilde{\tilde{g}}^{AB}
=\lambda \left({ l}^{AB}-\frac12 l \tilde{\tilde{g}}^{AB}\right)
 \, .
\end{equation}
The above results may be gathered as follows:
\begin{widetext}
\begin{align*}
2\dot\lambda\delta\alpha =&  \frac
{\sqrt{1+q^2}}{q}\left(2\dot\lambda
\delta {\mathfrak a} - 2 \dot\lambda \delta \log n\right) , \\
-2\dot\alpha\delta\lambda  = & \frac
{\sqrt{1+q^2}}{q}\left(-2\partial_0{\mathfrak a}\delta \lambda + 2
\partial_0 \log n \delta\lambda\right) ,\\ 2n\delta {\cal
Q}  = & 2\delta(\lambda l)-2\lambda l\delta(\log n) ,\\
-2n^A\delta {\cal  Q}_A  = & -2 n^A\delta(\lambda {\mathfrak
w}_A) - 2n^A \delta (\lambda(\log n )_{,A} ),\\
\stackrel{\perp}{{\cal  Q}}\!{^{AB}}g_{AB} \, \delta \log\lambda =
& -\left[ l +2
 {\mathfrak w}_0 - 2 n^A {\mathfrak w}_A + 2(\partial_0 -n^A
\partial_A)(\log n )\right]\delta\lambda , \\
\left(\stackrel{\perp}{{\cal  Q}}\!^{AB}-\frac12
\stackrel{\perp}{{\cal
Q}}\!{^{CD}}g_{CD}\tilde{\tilde{g}}^{AB}\right)\delta g_{AB}
 = & \lambda \left({ l}^{AB}-\frac12 l \tilde{\tilde{g}}^{AB}\right)
 \delta g_{AB}.
\end{align*}
\end{widetext} Finally, we express quantities appearing in boundary
formula (\ref{homog-form}) in terms of the above objects. Taking
into account identity (\ref{ldotjj})
and omitting two-dimensional divergencies we obtain
\begin{subequations}\label{homog-form2}
\begin{align}
0=\int_{V}&\left(  \dot{P}^{kl}\delta g_{kl} -\dot{g}_{kl}\delta
P^{kl}\right)\\&+2\int_{\partial V}
\frac{\sqrt{1+q^2}}{q}(\dot{\lambda}\delta {\mathfrak
a}-\dot{{\mathfrak a}}\delta\lambda)+2\delta\int_{\partial
V}\lambda l\\& + \int_{\partial V} (\lambda {l}^{AB} \delta g_{AB}
-2\lambda n^A\delta {\mathfrak w}_A -2{\mathfrak w}_0
\delta\lambda )\\&+2 \int_{\partial V}\left(\dot{\lambda}\delta
\log n- \frac {\sqrt{1+q^2}}{q}\dot{\lambda}\delta \log
n\right)\label{qln1}\\&+2\int_{\partial V}\left(\frac
{\sqrt{1+q^2}}{q}(\log n)^{\dot{}}\delta \lambda -(\log
n)^{\dot{}}\delta \lambda\right) . \label{qln}
\end{align}
\end{subequations}
In the null limit  $\epsilon\rightarrow 0$, we have that
${\mathfrak a}= a$ and ${\mathfrak w}_a = w_a$. Moreover, $q
\rightarrow \infty$ and, whence, ${\sqrt{1+q^2}}/q \rightarrow 1$.
We shall prove that the last two boundary terms (\ref{qln1}) and
(\ref{qln}) vanish in this limit. Indeed, we have
\be\label{limqdn} \lim_{\epsilon
 \rightarrow 0}
 \left(\frac {\sqrt{1+q^2}}{q} -1\right)
 \delta \log n = \frac{sM}{2N}\delta\sqrt{|g^{33}|} \, .
\ee A similar formula with $\delta$ replaced by $\partial_0$ is
also true. But $\delta g^{33}\rightarrow 0$ and
$\dot{g}^{33}\rightarrow 0$ on $\partial V$. This implies:
\begin{align*}
 \left. \lim_{\epsilon \rightarrow 0}
 \left(\frac {\sqrt{1+q^2}}{q} -1\right)
 \delta \log n \right|_{\partial V} =& 0,
 \\  \left. \lim_{\epsilon \rightarrow 0}
 \left(\frac {\sqrt{1+q^2}}{q} -1\right)
 \partial_0(\log n) \right|_{\partial V}=&0 .
\end{align*}
Hence, boundary formula (\ref{homog-form2}) takes the following
form:
\begin{align}
0
=&\int_{V}\left(  \dot{P}^{kl}\delta g_{kl}-\dot{g}%
_{kl}\delta P^{kl}\right)\nonumber\\&+2\int_{\partial V}
(\dot{\lambda}\delta {a}-\dot{ a}\delta\lambda)
+2\delta\int_{\partial V}\lambda l\nonumber\\&+ \int_{\partial V}
(\lambda { l}^{AB} \delta g_{AB} -2\lambda n^A\delta w_A -2w_0
\delta\lambda ).
\end{align}
Moving $2\delta\int_{\partial V}\lambda l$ to the left-hand side,
we obtain formula (\ref{form-zerowa1a}).

\section{Derivation of the formula (\ref{Omega1})}\label{omega-proof}

Being a difference between two symmetry fields: the
four-dimensional Killing field $\partial_0$ and the null field $K$
on $S$, the field $\vec{n}=(n^A)$ is a symmetry field of the
two-metric $g_{AB}$ and, whence, satisfies the two-dimensional
Killing equation:
\begin{equation}
n_{A\parallel B} + n_{B\parallel A}=0.
\end{equation}
As was already discussed,  we may choose a coordinate system such
that  $g_{AB}=f h_{AB}$, and $h_{AB}$ is a standard two-sphere
metric. The field $\vec{n}$ is also the symmetry field of
conformal structure given by the metric  $h_{AB}$, hence it has to
fulfill the equation:
\begin{equation}
n_{A\tilde{\parallel} B} + n_{B\tilde{\parallel}
A}-h_{AB}n^C{_{\tilde{\parallel} C}}=0,
\end{equation}
where $\tilde{\parallel}$ denotes the two-dimensional derivative
with respect to metric $h$ on the two-sphere. Therefore,
$\vec{n}$ belongs to the six-dimensional space of conformal fields
on two-sphere and it is of the following form:
\begin{equation}
n^A=\varepsilon^{AB}\stackrel{1}{v}_{,B}+\stackrel{2}{v}_{,B}h^{AB},
\end{equation}
where $\stackrel{1}{v}$,  $\stackrel{2}{v}$ are dipole functions,
i.e., of the form: $\stackrel{1}{v}=a_i k^i$, $\stackrel{2}{v}=b_i
k^i$; where $k^i$ are coordinates of the unit vector:
\begin{align*}
k^1&=\sin\theta\cos\varphi,\\ k^2&=\sin\theta\sin\varphi,\\
k^3&=\cos\theta.
\end{align*}
Taking all this into account we write the field $\vec{n}$ in the
following form:
\begin{align*}
n^A\partial_A&=\left(\varepsilon^{\theta\varphi}(a_i
k^i)_{,\varphi}
+(b_ik^i)_{,\theta}\right)\partial_{\theta}\nonumber\\&+
\left(\varepsilon^{\varphi\theta}(a_i
k^i)_{,\theta}+(b_ik^i)_{,\varphi}\frac1{\sin2\theta}\right)
\partial_\varphi.
\end{align*}
We will show that there exists a coordinate system in which
 $\vec{n}$ may be written as
\[
n^A\partial_A=-\Omega\frac{\partial}{\partial \varphi}.
\]
\begin{proof}
Suppose that for  $\theta=0$ component  $n^\theta$ vanishes. Hence
$a_1=b_2$ i $a_2=-b_1$. Then the components of $\vec{n}$ read as
\begin{align*}
n^\theta &=(\cos\theta-1)
(a_1\sin\varphi+b_1\cos\varphi)-b_3\sin\theta,\\
n^\varphi&=(\cos\theta-1)(b_1\sin\varphi-a_1\cos\varphi)
\frac{1}{\sin\theta}+a_3.
\end{align*}
We can rotate the coordinate system in  such a way that the
components $n^\theta $ and $n^\varphi $ will take the form:
\begin{align*}
n^\theta &=(\cos\theta-1)a\sin\varphi-b\sin\theta,\\ n^\varphi
&=(\cos\theta-1)(-a \cos\varphi) \frac{1}{\sin\theta}+c,
\end{align*}
$a$, $b$, $c$ are some new parameters. Because $S$ is a horizon,
we have:
\[
(\lambda n^A)_{,A}=0,
\]
and in our coordinate system $\lambda=f\sin\theta$, hence we have
the following equation:
\begin{align}\label{rownanie-z-c}
0&=\left\{ f\sin\theta\left(
(\cos\theta-1)a\sin\varphi-b\sin\theta
\right)\right\}_{,\theta}\nonumber\\ &+ \left\{ f\sin\theta \left(
(\cos\theta-1)(-a \cos\varphi) \frac1{\sin\theta}+c
\right)\right\}_{,\varphi}.
\end{align}
Integrating the above equation over  $\varphi$ and omitting
vanishing integral $\int (\lambda n^\varphi)_{,\varphi} d\varphi$
we obtain
\begin{equation}
\frac{\partial}{\partial\theta} \int_0^{2\pi} f\sin\theta\left(
(\cos\theta-1)a\sin\varphi-b\sin\theta \right) d\varphi=0,
\end{equation}
hence
\begin{equation}
\frac{\cos\theta-1}{\sin\theta}a \int_0^{2\pi}f\sin\varphi
d\varphi+b\int_0^{2\pi}f d\varphi=0.
\end{equation}
Because the first summand vanishes for $\theta\rightarrow 0$, we
have:
\[
2\pi b\ f(\theta=0)=b\int_0^{2\pi}f d\varphi=0\, .
\]
This straightforwardly implies $b=0$, because the conformal factor
$f$ is positive. Hence, $\vec{n}$ is of the following form:
\begin{align*}
n^\theta&=(\cos\theta-1)a\sin\varphi,\\
n^\varphi&=(\cos\theta-1)(-a \cos\varphi) \frac1{\sin\theta}+c.
\end{align*}
Let us write $\vec{n}$ in stereographic coordinates on the plane
$(x, y)$ intersecting our two-sphere along equator, where
coordinates $(x,y)$ read as:
\begin{align}
x&={\rm ctg}\frac{\theta}2\cos\varphi ,\nonumber\\ y&={\rm
ctg}\frac{\theta}2\sin\varphi.  \label{zmienne-st}
\end{align}
Then:
\begin{align*}
\frac{\partial}{\partial\theta}&=-\frac12
\frac{\cos\varphi}{\sin^2\frac{\theta}2}\frac{\partial}{\partial
x}-\frac12
\frac{\sin\varphi}{\sin^2\frac{\theta}2}\frac{\partial}{\partial
y}, \\ \frac{\partial}{\partial\varphi}&=-{\rm
ctg}\frac{\theta}2\sin\varphi\frac{\partial}{\partial x}+ {\rm
ctg}\frac{\theta}2\cos\varphi\frac{\partial}{\partial y},
\end{align*}
and $n^A\partial_A$ is of the following form:
\begin{equation}
n^A {\partial_A}= a\frac{\partial}{\partial y}
+c\left(-y\frac{\partial}{\partial x}+x\frac{\partial}{\partial
y}\right).
\end{equation}
Let us consider two cases: $c=0$ and $c\neq0$:
\begin{itemize}
\item{\bf $c=0$}\\ In this case equation (\ref{rownanie-z-c})
takes the following form:
\begin{align}\label{rownanie-z-c1}
&\left( f\sin\theta(\cos\theta-1)a\sin\varphi
\right)_{,\theta}\nonumber\\ &\qquad+ \left( f\sin\theta
(\cos\theta-1)(-a \cos\varphi) \frac1{\sin\theta}
\right)_{,\varphi}=0.
\end{align}
In stereographic coordinates  (\ref{zmienne-st}) the above
equation reads:
\begin{equation}\left(\log f\right)_{,y}=\frac{2y}{1+x^2+y^2},
\end{equation}
and implies
\begin{equation}
f(x,y)=C(x)(1+x^2+y^2),
\end{equation}
where $C(x)$ is an arbitrary function of $x$. Keep  $x$ constant
and pass to the limit $y \rightarrow \infty$. If  $C(x)\neq 0$
then $f\rightarrow\infty$, otherwise $f\equiv 0$. But the
conformal factor $f$ must be finite and different from zero.
Hence, the case $c=0$ is incompatible with the properties of $f$.
\item{\bf $c\neq0$}\\
 In that case we may write
\begin{equation}
n^A\partial_A =
c\left[\left(x+\frac{a}c\right)\frac{\partial}{\partial
y}-y\frac{\partial}{\partial x}\right].
\end{equation}
Putting $\tilde{x}=x+\frac{a}c$, we obtain:
\begin{equation}
n^A{\partial_A}= c\left(\tilde{x}\frac{\partial}{\partial
{y}}-{y}\frac{\partial}{\partial \tilde{x}}\right),
\end{equation}
hence $n^A\partial_A$ is a field of rotations and can be written
as follows:
\begin{equation}\label{obrot}
n^A\partial_A=-\Omega\frac{\partial}{\partial \tilde\varphi}.
\end{equation}
\end{itemize}
\end{proof}
A different proof of this fact is given in the paper
\cite{lewandowski}.

\section{Surface gravity $\kappa$ is constant
along $\vec{n}$} \label{kappa-proof}

We are going to perform a gauge transformation of $\kappa$, such
that the resulting quantity $\tilde\kappa$ remains constant along
$\vec{n}$. For this purpose we leave the spacelike coordinates
unchanged ($\tilde{x}^k = x^k$), whereas the time coordinate is
translated by a constant which depends upon them. This implies the
following transformation on the surface $S$, where $x^3$ is
constant:
\begin{equation}\label{trans}
  {\tilde x}^0 = x^0 + \alpha(x^A) .
\end{equation}
Then the field $K$ transforms as ${\tilde K} = c K$, where
\begin{equation}\label{trans1}
c = \left( 1 - n^A \partial_A \alpha \right)^{-1} ,
\end{equation}
and quantities $w_a$
 transform as follows:
\begin{equation}\label{nowe-w}
  {\tilde w}_a=w_a+\partial_a(\log c)
\end{equation}
(transformation law for objects $w_a$ is given in \cite{JKC}).
After this transformation the surface gravity ${\tilde \kappa}$
takes the following value:
\begin{equation}\label{w-tilde}
  {\tilde \kappa} =- {\tilde K}^a  {\tilde w}_a=
  -cK^a\left(w_a+\frac{\partial_ac}{c}\right)=c\kappa-K(c).
\end{equation}
Using $K=\partial_0-n^A\partial_A$ and formula \eqref{obrot} for
$n^A$ we obtain
\begin{equation}
\Omega\frac{\partial c}{\partial\varphi}-c\kappa+{\tilde\kappa}=0
.
\end{equation}
Consequently,
\begin{equation}
c=-\int\frac{\tilde\kappa}{\Omega}e^{(-\Omega^{-1}\int\kappa
d\varphi)}d\varphi.
\end{equation}
Denoting
\begin{equation}
F(\theta,\varphi)=\int e^{(-\Omega^{-1}\int\kappa
d\varphi)}d\varphi
\end{equation}
we obtain
\begin{equation}
c^{-1}=-\frac{\Omega}{\tilde
\kappa}\frac{\partial}{\partial\varphi}\left(\log F\right).
\end{equation}
We can compare this with the form of $c^{-1}$ implied by
(\ref{trans1})
\begin{equation}
c^{-1}=1-\frac{1}{\Omega\frac{\partial\alpha}{\partial\varphi}},
\end{equation}
which leads to the following equation:
\begin{equation}
\frac{\partial\alpha}{\partial\varphi}=\frac1\Omega+
\frac1{\tilde\kappa}\frac{\partial}{\partial\varphi}\log F.
\end{equation}
Solutions of the above equation reads as
\begin{equation}
\alpha=\frac{\varphi}{\Omega}+\frac1{\tilde\kappa}\log F,
\end{equation}
where
\begin{equation*}
F(\varphi)=C_1\left( \int_0^\varphi
e^{(-\Omega^{-1}\int_0^u\kappa(s) ds)}du+ C_2 \right),
\end{equation*}
and $C_1$, $C_2$ are integration constants. Applying periodicity
conditions:
\begin{align*}
\alpha(0)&=\alpha(2\pi),\\ \frac{d}{d\varphi}(\log F)(0)
&=\frac{d}{d\varphi}(\log F)(2\pi)
\end{align*}
we obtain
\begin{equation}
{\tilde\kappa}=\frac{\Omega}{2\pi}\log\frac{F(0)}{F(2\pi)},
\end{equation}
where the  values of $F(0)$ and $F(2\pi)$ remain to be determined.
Denote by $f(\varphi)$ the expression:
\begin{equation}
f(\varphi):=\int_0^\varphi e^{(-\Omega^{-1}\int_0^u\kappa(s)
ds)}du .
\end{equation}
Therefore $F$ and $\log F$ are of the following form:
\begin{align*}
F&=C_1(f+C_2),\\ \log F&=\log C_1+ \log(f+C_2).
\end{align*}
The above equations imply
\begin{equation}
\frac{f'(0)}{f(0)+C_3} = \frac{f'(2\pi)}{f(2\pi)+C_3}\ \ {\rm
and}\ \ f(0)=1,
\end{equation}
hence \begin{equation}  C_3 = \frac{f(2\pi)}{f'(2\pi)-1},
\end{equation}
and we have
\begin{equation}
 {\tilde\kappa}=\frac{\Omega}{2\pi}
\log\frac{C_3}{f(2\pi)+C_3}.
\end{equation}
Finally, we obtain
\begin{equation}
{\tilde\kappa}=-\frac{\Omega}{2\pi} \log
f'(2\pi)=\frac1{2\pi}\int_0^{2\pi} \kappa(s)ds\equiv
\textrm{const}.
\end{equation}
Hence, we have performed such a gauge transformations from
$\kappa$ to $\tilde\kappa$, that the new quantity $\tilde{\kappa}$
is constant along parallels of the sphere $S^2$.


\end{document}